\documentclass[11pt]{article}
\usepackage{amsmath, amsfonts, url}
\usepackage[T1]{fontenc}
\usepackage[latin1]{inputenc}
\usepackage{amssymb}

\begin{document}

\makeatletter

\newenvironment{algorithm}{\begin{algorithm1}\ \\
    \vspace{-0.2cm}}{\end{algorithm1}}

\newenvironment{proofsk}{\begin{proof}[Proof Sketch:]}
{\end{proof}}

\newenvironment{smallproof}{\nopagebreak \begin{quote} %
\begin{small} \noindent{\bf Proof:}}{ \qed \par %
\end{small} \end{quote} \medskip}

\newenvironment{note}{\nopagebreak \begin{quote} %
\noindent{\bf Note:}}{%
\end{quote} \medskip}

\newenvironment{notes}{\nopagebreak \begin{quote} %
\noindent{\bf Notes:} \par%
\begin{itemize}}{%
\end{itemize}\end{quote} \medskip}

\newenvironment{summary}{\begin{quote} {\bf Summary:}}{\end{quote}}


\newcommand{\eqdef}{\stackrel{def}{=}}
\newcommand{\N}{\mathbb{N}}
\newcommand{\R}{\mathbb{R}}
\newcommand{\C}{\mathbb{C}}
\newcommand{\Z}{\mathbb{Z}}
\newcommand{\F}{\mathbb{F}}
\newcommand{\Zn}{{\Z}_n}
\newcommand{\bits}{\{0,1\}}
\newcommand{\inr}{\in_{\mbox{\tiny R}}}
\newcommand{\getsr}{\gets_{\mbox{\tiny R}}}
\newcommand{\st}{\mbox{ s.t. }}
\newcommand{\etal}{{\it et al }}
\newcommand{\into}{\rightarrow}

\newcommand{\Ex}{\mathbb{E}}
\newcommand{\To}{\rightarrow}
\newcommand{\e}{\epsilon}
\newcommand{\ee}{\varepsilon}
\newcommand{\ceil}[1]{{\lceil{#1}\rceil}}
\newcommand{\floor}[1]{{\lfloor{#1}\rfloor}}
\newcommand{\angles}[1]{\langle #1 \rangle}
\newcommand{\var}{\mbox{var}}
\newcommand{\trace}{\mbox{trace}}
\newcommand{\ignore}[1]{}
\newcommand{\Alg}{\mathrm{Alg}}
\newcommand{\fro}[1]{\|#1\|_F}
\newcommand{\trn}[1]{\|#1\|_{tr}}
\newcommand{\norm}[1]{\|#1\|}

\newcommand{\NP}{\mathbf{NP}}
\renewcommand{\P}{\mathbf{P}}
\newcommand{\PCP}{\mathbf{PCP}}
\newcommand{\RP}{\mathbf{RP}}
\newcommand{\BPP}{\mathbf{BPP}}

\newcommand{\Tr}{\mathrm{Tr}}

\newcommand{\lang}{\mathcal{L}}

\newcommand{\Poincare}{Poincar�}


\newtheorem{theorem}{Theorem}
\newtheorem{lemma}[theorem]{Lemma}
\newtheorem{claim}[theorem]{Claim}
\newtheorem{corollary}[theorem]{Corollary}
\newtheorem{conjecture}[theorem]{Conjecture}
\newtheorem{question}[theorem]{Question}
\newtheorem{proposition}[theorem]{Proposition}
\newtheorem{axiom}[theorem]{Axiom}
\newtheorem{remark}[theorem]{Remark}
\newtheorem{example}[theorem]{Example}
\newtheorem{exercise}[theorem]{Exercise}
\newtheorem{definition}[theorem]{Definition}
\newtheorem{observation}[theorem]{Observation}

\def\pproof{\par\penalty-1000\vskip .5 pt\noindent{\bf Proof\/ }}
\newcommand{\QED}{\hfill$\;\;\;\rule[0.1mm]{2mm}{2mm}$}

\newenvironment{proof}{\begin{pproof}}{\QED\end{pproof}~\\}


\newcommand{\NChooseM}[2]{\ensuremath{\lp{(}\begin{array}{cc}#1\\#2\end{array}\rp{)}}}
\newcommand{\mc}[1]{{\cal{#1}}}
\newcommand{\lp}[1]{\left #1}
\newcommand{\rp}[1]{\right #1}
\newcommand{\vect}[1]{\ensuremath{{\mathbf #1}}}
\newcommand{\IP}[2]{\ensuremath{\lp{<}#1,#2\rp{>}}}
\newcommand{\sv}{\ensuremath{\vect{v}}}
\newcommand{\sve}{\ensuremath{v}}

\newcommand{\op}{W}
\newcommand{\opr}{\vect{w}}
\newcommand{\allvects}[1]{L_{#1}}
\newcommand{\avn}[1]{L_{#1}}
\newcommand{\av}{L_n}
\newcommand{\ff}{\lfloor f \rfloor}

\newcommand{\sizeo}[1][\e]{\mathrm{size}^{(v)}_{#1}}
\newcommand{\corro}[2][\e]{\mathrm{ubc}^{(#2)}_{#1}}
\newcommand{\corr}[1][\e]{\mathrm{ubc}_{#1}}
\newcommand{\rcorro}[2][\e]{\mathrm{corr}^{(#2)}_{#1}}
\newcommand{\rcorr}[1][\e]{\mathrm{corr}_{#1}}
\newcommand{\size}[1][\e]{\mathrm{size}_{#1}}
\newcommand{\scorr}[1][\e]{\mathrm{sc}_{#1}}
\newcommand{\monoo}[2][\rho]{\mathrm{mono}^{(#2)}_{#1}}
\newcommand{\mono}[1][\rho]{\mathrm{mono}_{#1}}
\newcommand{\hmonoo}[2][\rho]{\mathrm{hmono}^{(#2)}_{#1}}
\newcommand{\hmono}[1][\rho]{\mathrm{hmono}_{#1}}

\newcommand{\rk}{\mathrm{rank}}

\title{The Corruption Bound, Log-Rank, and Communication Complexity}

\author{Adi Shraibman \\ The School of Computer Science \\  
The Academic College of Tel Aviv-Yaffo\\{\tt adish@mta.ac.il}}

\date{}

\maketitle

\abstract{
We prove upper bounds on deterministic communication complexity in terms of log of the rank
and simple versions of the corruption bound. 

Our bounds are a simplified version of the results of Gavinsky and Lovett \cite{LG13}, using the same set of tools. 
We also give an elementary proof for the upper bound on communication complexity in terms of rank proved by 
Lovett \cite{Lov13}.
}

\section{Introduction}

\subsection{An overview}

The notion of communication complexity was introduced by Yao \cite{Yao79} as a discrete variation
of a model of Abelson \cite{Abel} concerning information transfer in distributive computations.
In the basic model, two players Alice and Bob wish to compute together a discrete boolean function 
$f:I \times J \to \{0,1\}$. The players are assumed to have infinite computational power, and their
goal is to minimize the communication between them during the computation.  
Alice and Bob first agree on a communication protocol and then use this 
protocol to compute the value of the function $f$ on any given pair of feasible inputs $(i, j) \in I\times J$.
The inputs are presented to the players in a way that Alice sees only $i$ and Bob receives $j$. They then take turns writing
bits ($0$ or $1$) on a blackboard, until both players know the value of $f(i,j)$.
The {\em cost} of a protocol is the maximal number
of bits the players write on the blackboard during the computation of $f(i,j)$, over
all choices of inputs $(i, j) \in I\times J$. The {\em deterministic communication complexity}
of $f$, denoted $D(f)$, is equal to the minimal cost of a protocol for $f$.

It is sometimes convenient to consider the function $f:I \times J \to \{0,1\}$ as a sign matrix $A$, 
where the rows of $A$ correspond to $i \in I$ and the columns correspond to $j \in J$. The entries of
$A$ satisfy $A_{i,j} = 1$ if $f(i,j) = 0$ and $A_{i,j} = -1$ if $f(i,j) = 1$. We use this matrix notation.

Many variants of the basic communication complexity model of Yao were defined. The models
differ in the type of communication allowed (e.g. deterministic, randomized, nondeterministic, etc),
the number of players, the type of function computed, and more. The interested reader can see 
the book of Kushilevitz and Nisan \cite{KN97} for a thorough exposition and discussion on 
the basic models of communication complexity.

The communication complexity literature is mainly concerned with proving lower bounds,
and indeed communication complexity lower bounds are used in various areas of theoretical
computer science such as decision tree complexity, VLSI circuits, space-time tradeoffs for 
Turing machines and more.
To prove lower bounds in communication complexity many techniques were developed.
One of the early lower bound techniques is the rank lower bound of Mehlhorn and Schmidt \cite{MS82}:
Let $A$ be a sign matrix, and denote by $\rk(A)$ the rank of $A$ over the reals,
then $$\log_2 \rk(A) \le D(A).$$
Since there are many variants of communication complexity and various lower bounds, it is interesting
to fully understand the relation between the different measures of complexity. For this purpose we need to prove 
upper bounds as well as lower bounds. 

Some of the major open questions in communication complexity are of 
this type, and in particular the log-rank conjecture.
The log-rank conjecture \cite{LS93, NW95} states that deterministic communication complexity of a sign matrix $A$ and 
log of the rank of $A$ are polynomially related. Namely, the conjecture is that there is a constant $c$
such that every sign matrix $A$ satisfies $D(A) \le (\log \rk(A))^c$. A fairly simple upper bound is
$D(A) \le\rk(A)$, and that was the best known for quite a while. Drawing ideas and tools from \cite{BLZ12,LG13,LMSS07,LS08b}, Lovett 
\cite{Lov13} proved the bound $D(A) \le O(\sqrt{\rk(A)} \log \rk(A))$, achieving a first significant breakthrough at this end.  On the other end, it is known that
the constant $c$ in the log-rank conjecture must be at least two \cite{goos2015deterministic}. This is
also a recent breakthrough, continuing a line of work \cite{AS89, Raz92b, RS93, NW95} which gradually 
constructed matrices with larger gaps between the log of the rank and the deterministic 
communication complexity.

Although the gap in our knowledge regarding the log-rank conjecture is very wide, there are interesting
upper bounds on $D(A)$, which include:
\begin{enumerate}

\item $D(A) \le (N^0(A) +1 )(N^1(A)+1)$ \cite{AUY83},

\item $D(A) \le \log \rk(A) \log \rk_+(A)$ \cite{Lov90},

\item $D(A) \le \log (\rk(A)+1) (\min\{N^0(A),N^1(A)\}+1)$ \cite{Lov90},

\item $D(A) \le fs(A) (N^0(A)+1)$ \cite{Lov90}.

\end{enumerate}
Here, $N^1(A)$ and $N^0(A)$ are the nondeterministic and co-nondeterministic communication complexity of $A$ respectively, 
$\rk_+(A)$ is the positive rank of $A$, and $fs(A)$ is the maximal size of a fooling set for $A$. 
All these complexity measures are known lower bounds on deterministic communication complexity. 
See \cite{Lov90} and also \cite{KN97} for a comprehensive survey of these complexity measures and bounds.
See \cite{dietzfelbinger1996comparison} for a comparison between $fs(A)$ and $rank(A)$, 
and also an extension 
of the fooling set bound which is polynomially tight up to a logarithmic additive factor.

In the above bounds, roughly speaking, $fs(A)$ or $\rk(A)$ serve as a potential
function, while $N^0(A)$ or $N^1(A)$ serve as a pool of monochromatic rectangles. 
Another upper bound, similar in nature, is given by Nisan and Wigderson \cite{NW95}.
The statement, as it was phrased in \cite{LG13}, is
\begin{theorem}[\cite{NW95, LG13, Lov13}]
\label{claim:gen_bound}
Let $A$ be a sign matrix and let $\rk(A)=r$. Assume that every submatrix $B$ of $A$
contains a monochromatic rectangle of size at least $2^{-q}|B|$, where $|B|$ is the
size of $B$ (i.e., the number of entries). Then, 
$$
D(A) \le O(\log^2 r+q\log r).
$$ 
\end{theorem}
In the protocol of Nisan and Wigderson, rank serves like before as a kind of a potential
function. $N^0(A)$ and $N^1(A)$ on the other hand are replaced by 
the size of a largest monochromatic rectangle which appears in their bound as an
independent complexity measure.

Gavinsky and Lovett \cite{LG13} augmented the above repertoire of upper bounds on deterministic
communication complexity. They proved that the 
deterministic communication complexity of a sign matrix $A$ is at most $O(CC(A) \log^2 \rk(A))$, where $CC(A)$ is either 
randomized communication complexity, information complexity, or zero-communication complexity. 
Thus when the rank of the matrix is low, an efficient nondeterministic protocol or a randomized protocol, implies an efficient 
deterministic protocol.

The core upper bound of \cite{LG13} is in terms of extended discrepancy and it therefore implies additional results to those listed above. 
In fact, as observed by G\"o\"os and Watson \cite{GW14}, extended discrepancy corresponds to a fractional version of approximate majority covers (see \cite{Kla03} for details).
Thus, for example, the bounds of \cite{LG13} are also valid for the Merlin Arthur (MA) complexity of $A$, with error $1/4$. In this model the players first make a nondeterministic guess and then perform a randomized protocol. This has the nice interpretation that when the rank is low, there is an efficient deterministic protocol, even compared with protocols combining the power of nondeterminism and randomization.

The proofs of \cite{LG13} are based on the protocol of Nisan and Wigderson \cite{NW95}
(Theorem~\ref{claim:gen_bound}). In addition they use a simple and clever lemma relating the size of almost 
monochromatic rectangles, in some conditions, to the size of monochromatic ones.
\begin{theorem}[\cite{LG13, Lov13}]
\label{claim:corr_to_mono}
Let $A$ be an $m \times n$ sign matrix with $\rk(A)=r$. Assume that the fraction of $1$'s or the fraction of $-1$'s in
$A$ is at most $\frac{1}{4r}$. Then $A$ contains a monochromatic rectangle $R$ such that $|R| \ge \frac{mn}{8}$.  
\end{theorem}
Theorem~\ref{claim:gen_bound} gives an upper bound on $D(A)$ in terms of the size
of monochromatic rectangles. Theorem~\ref{claim:corr_to_mono} enables to replace the concept
of monochromatic rectangles in this bound by a more relaxed notion.

Our contribution is to point out that the language of corruption bounds is perfectly suited for the line 
of proof described above. This enables to give a simple and elementary proofs for the results of 
\cite{LG13} and \cite{Lov13}. Furthermore, the corruption bound, which is a central lower bound 
technique in (randomized) communication complexity, has proved relations with: 
randomized communication complexity, information complexity, zero-communication complexity,
nondeterministic communication complexity, MA complexity,
and positive rank. The corruption bound therefore provides
a uniform view of previous upper bounds, as well as a natural link to the upper bound of Nisan and Wigderson
which is based on the size of monochromatic rectangles.

\subsection{The corruption bound}
\label{definitions}

The heart of the matter is the following definition:
\begin{definition}
Let $A$ be an $m \times n$ sign matrix, and 
denote by $u$ the uniform distribution on $[m]\times [n]$.
Let $v \in \{-1,1\}$ be such that $u(v) \le 1/2$.\footnote{If $u(1)=u(-1)$
then we let $v=-1$.}
For $0 \le \rho \le 1$ define 
$$
\mono(A) = \log \left( \frac{1}{\max_R \{u(R): u(v|R) \le \rho u(v)\}} \right) , 
$$
where the maximum is over all combinatorial rectangles $R$. 
Let $\hmono(A)$ be the maximum of $\mono(B)$ over all submatrices $B$ of $A$.
\end{definition}

Denote the distribution of $-1$'s and $1$'s in $A$ by $(\alpha,1-\alpha)$, and assume without loss of generality
that $\alpha \le 1/2$. Then roughly speaking, $\mono(A)$ quantifies the relative size of a largest submatrix of $A$ in which
the frequency changes to $(\rho \alpha, 1- \rho \alpha)$ or even more biased. That is, we seek a submatrix
for which the frequency of $-1$'s is smaller by at least a factor of $\rho$ than their frequency in $A$.
In particular, $\mono[0](A)$ quantifies the relative size of a largest submatrix of $A$ containing only
$1$-entries. 

The quantity $\hmono(A)$ is a hereditary version of $\mono(A)$, that is obviously needed 
if we wish to show a strong relation with
communication complexity which is hereditary.  
It is an easy exercise to show that $\hmono[0](A) \le D(A)+1$
\footnote{This inequality follows from the fact that every $c$-bit deterministic
communication protocol for $A$ partitions the matrix $A$ into at most $2^c$
monochromatic combinatorial rectangle.}
, and the protocol of 
Nisan and Wigderson \cite{NW95} implies that $D(A) \le O(\log^2 r + \hmono[0](A) \log r)$.
Therefore, the log-rank conjecture is true if and only if $\hmono[0](A) \le (\log \rk(A))^c$ for some constant $c$.
This kind of relation though (if true), between the rank of a matrix and the size of a monochromatic submatrix,
is very hard to capture. The contribution of Theorem~\ref{claim:corr_to_mono} is that instead of monochromatic
submatrices we can consider the more relaxed notion $\hmono[1/2](A)$, that is:
\begin{theorem}
\label{main_bound_on_cc}
For every sign matrix $A$ with $r = rank(A)$ it holds that $$D(A) \le O(\hmono[1/2](A)\log^2 r).$$ 
\end{theorem}

Clearly, $\mono[\rho_1](A) \le \mono[\rho_2](A)$ whenever $\rho_1 \ge \rho_2$, thus the above
upper bound via $\hmono[1/2](A)$ is tighter than the previous bound in terms of $\hmono[0](A)$
when ignoring the log-rank factors. The more important advantage is though that the nature of $\hmono[1/2](A)$
makes it easier to relate it to other complexity measures such as randomized communication complexity, 
information complexity, zero-communication complexity, and more \cite{LG13}. This enhances the variety 
of upper bounds on deterministic communication complexity that are applicable when the rank is small.
All these upper bounds follow from the relation between $\hmono[1/2](A)$ and the corruption
bound, explained in Section~\ref{sec:relation_cor_bnd}.

Another complexity measure of a sign matrix $A$ that can be tied to  $\hmono[1/2](A)$ is the discrepancy of $A$, denoted
$disc(A)$, which is defined as follows: Let $\sigma$ 
be a distribution on the entries of $A$. The discrepancy with respect to $\sigma$
is:
$$
\max_R |\sigma(1,R)-\sigma(-1,R)|,
$$ 
where the maximum is 
over all combinatorial rectangles in $A$. The discrepancy of $A$ is the minimal 
discrepancy with respect to $\sigma$, over all probability distributions.

The discrepancy is often used to lower bound communication complexity 
in different models, and it is also equivalent (up to a constant) to the reciprocal of margin complexity. 
See \cite{LMSS07, LS08b} for the definitions and proof of the equivalence of 
these measures. This equivalence was used in \cite{LS08b} to prove that 
$1/disc(A) \le O(\sqrt{\rk(A)})$.

For a sign matrix $A$ let $d=1/disc(A)$ and $r = rank(A)$. Lovett \cite{Lov13} proved that 
A contains a monochromatic rectangle of size $2^{-O(d \log r)}|A|$.
Combined with the protocol of Nisan and wigderson (Theorem~\ref{claim:gen_bound}) and the relation between discrepancy and rank,
this proves that deterministic communication complexity is bounded by root of the rank up to log factors. 
We give in Section~\ref{log-rank} an elementary proof
of a slightly different statement $\hmono[1/2](A) \le O(d \log d)$, that gives the same 
result up to a log factor.

\section{Notation}

Let $A$ be an $m \times n$ sign matrix, and let $\mu$ be a probability distribution on $[m]\times[n]$.
For a set of entries $E \subseteq [m]\times[n]$, let $\mu(E)$ be the sum $\sum_{(i,j) \in E} \mu(i,j)$.
A combinatorial rectangle is a subset $S\times T$ of entries, such that $S \subseteq [m]$ and $T \subseteq [n]$.
That is, a combinatorial rectangle corresponds to a submatrix of $A$.
With a slight abuse of notation, for $v \in \{\pm 1\}$ and a combinatorial rectangle $R$,
we denote by $\mu(v)=\mu(\{(i,j)|A_{i,j}=v\})$, and $\mu(v, R)=\mu(\{(i,j)\in R|A_{i,j}=v\})$.
We also write $\mu(v | R)$ for the probability that $A_{i,j}=v$ conditioned upon $v \in R$, which 
is equal to $\mu(v,R)/\mu(R)$. 
\newline
We call a distribution $\mu$ on $[m]\times[n]$ {\em uniformly-balanced} for $A$, if it satisfies:
\begin{itemize}

\item The set of entries $(i,j)$ for which $\mu(i,j) > 0$ is a combinatorial rectangle.

\item $A_{i,j} = A_{x,y}$ implies that $\mu(i,j) = \mu(x,y)$, if both are nonzero.

\item $\mu(1) = \mu(-1) = \frac{1}{2}$.

\end{itemize}

We use uniformly balanced distributions  in Section~\ref{sec:relation_cor_bnd} to define 
a simple version of the corruption bound and relate it to $\hmono(A)$.

\section{The upper bound}
\label{upper-bound}

To prove Theorem~\ref{main_bound_on_cc} we need a slight variation of Theorem~\ref{claim:corr_to_mono}.
\begin{claim}
\label{claim:corr_to_mono_variation}
Let $A$ be an $m \times n$ sign matrix with $\rk(A)=r$, and $v \in \{-1,1\}$. Assume that the fraction of $-v$'s 
in $A$ is at most $1/10r$. 
Then $A$ contains a $v$-monochromatic rectangle $R$ such that $|R| \ge \frac{mn}{8}$.  
\end{claim}

\begin{proof}
The proof follows from Theorem~\ref{claim:corr_to_mono} by observing that $R$ is too big to contain only $-v$ entries, as the fraction
of $-v$'s in $A$ is at most $1/10r \le 1/10$.
\end{proof}

\begin{proof}[of Theorem~\ref{main_bound_on_cc}]
First, recall that $D(A) \le O(\log^2 r + \hmono[0](A) \log r)$ by Theorem~\ref{claim:gen_bound}.
Second, the bound $\hmono[0](A) \le O(\hmono[1/10r](A))$ follows from Claim~\ref{claim:corr_to_mono_variation}.
Finally, observe that for every $0 \le \rho_1,\rho_2 \le 1$ it holds that $$\hmono[\rho_1\rho_2](A) \le \hmono[\rho_1](A) + \hmono[\rho_2](A).$$
The proof of the above inequality is straightforward from the definition. This makes amplification possible, and in particular shows that 
$\hmono[1/10r](A) \le O(\log r \cdot \hmono[1/2](A))$. Combining these inequalities gives the proof.
\end{proof}

\section{Relation with the corruption bound}
\label{sec:relation_cor_bnd}

We would like to show that $\hmono(A)$ is a lower bound for
randomized communication complexity, information complexity, zero-communication complexity,
nondeterministic communication complexity, and positive rank. The simplest way to do that is to relate
it to the corruption/rectangle bound 
(defined in the sequel) which was proved as a lower bound for all these complexity measures,
and more.

In a way $\hmono(A)$ is a very simple version of the corruption bound.
When Yao first used the corruption bound as a lower bound on randomized communication complexity, 
his definition was in the spirit of $\mono(A)$ (See Lemma~3 in \cite{Yao83}). This first bound did not
take into account the fact that one can choose any probability distribution over the entries of the matrix,
and not only the uniform distribution. It is easy to make $\mono(A)$ small even for matrices with high
randomized communication complexity, for example by planting a large monochromatic submatrix in
a random sign matrix. This can be fixed by considering the worst probability distribution
over the entries of the matrix, which can for example give zero weight to the large monochromatic
submatrix.
The corruption bound is usually defined as follows \cite{BFS86, Raz92d, Kla03, BPSW06, JK09-Journal}: 
\begin{definition}
Let $A$ be a sign matrix,  $0 \le \e \le 1$, and $v \in \{-1,1\}$. For a probability
distribution $\mu$ on the entries of $A$, define
$$
\sizeo(A, \mu) = \max_R \{\mu(R) : \mu(-v|R) \le \epsilon \},
$$
where the maximum is over all combinatorial rectangles $R$. Define
$$
\rcorro{v}(A) = \max_{\mu} \log 1/\sizeo(A, \mu),
$$
where $\mu$ runs over all balanced distributions
\footnote{A balanced distribution is a distribution for which the probability of $-1$'s 
and the probability of $1$'s are bounded from below by a constant.} 
for $A$. Finally define
$$
\rcorr(A) = \max \{ \rcorro{1}(A), \rcorro{-1}(A)\}.
$$
\end{definition}

We use a simple version of the corruption bound which is similar to the above, only 
we maximize over the family of uniformly-balanced distributions (defined in Section~\ref{definitions}), 
instead of all balanced distributions.
We denote this variant of the corruption bound by $\corr(A)$. Obviously $\corr(A) \le \rcorr(A)$ 
for every sign matrix $A$ and $0 \le \e \le 1$, as we maximize over a subfamily of probability
distributions.

As we show next, $\corr(A)$ is closely related to $\hmono[2\e](A)$. 
\begin{lemma}
\label{equiv_corr_hmono}
Let $A$ be an $m \times n$ sign matrix.
Then, for every $0 \le \e \le 1/2$ it holds that 
$$
\hmono[2\e](A) \le \corr(A).
$$
\end{lemma}

\begin{proof}
Let $u$ be the uniform distribution on $[m]\times [n]$, and
denote by $\mu$ the uniformly-balanced distribution for $A$ supported on $[m] \times [n]$.
For every entry $(i,j)$ it holds that
\begin{eqnarray}
\label{1}
A_{i,j} = -1 &\Rightarrow& \mu(i,j) = \frac{u(i,j)}{2u(-1)},\\ 
A_{i,j} = 1 &\Rightarrow&  \mu(i,j) = \frac{u(i,j)}{2u(1)}.  \nonumber
\end{eqnarray} 

The proof is very simple, but technical. We therefore first give the basic intuition. 
Consider the extreme case where 
$u(1)=u(-1)=1/2$. In this case $\mu = u$ and seeking a combinatorial rectangle
with $\mu(-1|R) \le \e$ and large $\mu(R)$ is equivalent to seeking a combinatorial
rectangle with $u(-1|R)\le 2\e u(-1)$ and large $u(R)$. Thus in this case the quantities
of interest both for $\corr(A)$ and for $\hmono[2\e](A)$ coincide. Since we consider 
uniformly balanced distributions, and normalize the distribution so that the probability 
of $1$'s and $-1$'s are equal,
the general case is similar. In the general case there is an advantage 
towards $\hmono[2\e](A)$ that increases when the imbalance between
the number of $1$'s and $-1$'s increases. The proof is essentially to show that the rectangle $R$ found
for $\corr(A)$ is also good for $\hmono[2\e](A)$. In the first part of 
the proof below we show that for this rectangle $u(R) \ge \mu(R)$, and in the second part
that $u(-1|R) \le 2 \e u(-1)$.

Assume without loss of generality that $u(-1) \le 1/2$, and let $\corr(A)=k$. 
Then, there is a rectangle $R$ such that $\mu(R) \ge 2^{-k}$ and $\mu(-1|R) \le \e$.  
Thus
\begin{eqnarray*}
u(R) &=& \sum_{(i,j) \in R} u(i,j)\\
&=& \sum_{A_{i,j}=-1, (i,j)\in R} u(i,j) + \sum_{A_{i,j}=1, (i,j)\in R} u(i,j) \\
&=& \sum_{A_{i,j}=-1, (i,j)\in R} 2u(-1)\mu(i,j) + \sum_{A_{i,j}=1, (i,j)\in R} 2u(1)\mu(i,j)\\
&=&  2u(-1)\mu(-1,R) + 2u(1)\mu(1,R)\\
&=& 2\mu(R)\left[ u(-1)\mu(-1|R) + \left(1-u(-1)\right)\left(1-\mu(-1|R)\right)\right]\\
&=& 2\mu(R)\left[ 1-u(-1)-\mu(-1|R) +2u(-1)\mu(-1|R)\right]\\
&\ge& \mu(R).
\end{eqnarray*}
The last step follows from the fact that the function 
$f(x,y)=1-x-y+2xy$ satisfies $f(x,y) \ge 1/2$ for $x,y \in [0,1/2]$. 
Recall that $u(-1) \le 1/2$ and $\mu(-1|R) \le \e \le 1/2$.
Also
\begin{eqnarray*}
u(-1,R) &=& \sum_{A_{i,j}=-1, (i,j) \in R} u(i,j)\\
&=& \sum_{A_{i,j}=-1, (i,j)\in R} 2u(-1)\mu(i,j)\\
&=&  2u(-1)\mu(-1,R)\\
&\le& 2 u(-1) \e \mu(R)\\
&\le& 2 \e u(-1) u(R).
\end{eqnarray*} 
This concludes that $\mono[2\e](A) \le k$ since $u(R) \ge \mu(R) \ge 2^{-k}$ and $u(-1|R) \le 2 \e u(-1)$. 
Proving similarly for every submatrix $B$ of $A$ gives $\hmono[2\e](A) \le k$.  
\end{proof}

\section{Proof of Theorem~\ref{cor-disc}}
\label{log-rank}

\begin{theorem}[\cite{Lov13}]
\label{cor-disc}
Let $A$ be a sign matrix, and let $disc(A)=1/d$. Then
$$
\hmono[1/2](A) \le O(d\log d).
$$
\end{theorem}

\begin{proof}
The proof is essentially observing that discrepancy corresponds to the error in corruption. 
Suppose that $\corr[1/2-1/6d](A) = O(\log d)$, then by Lemma~\ref{equiv_corr_hmono}
this implies that $\hmono[1-1/3d](A) \le O(\log d)$, and therefore
\[
\hmono[1/2](A) \le O(d\log d).
\]

We now prove that $\corr[1/2-1/6d](A) = O(\log d)$.
Let $\mu$ be a uniformly-balanced distribution for $A$.  
Since $disc(A)=disc(-A)$ it is enough to prove the existence 
of a large combinatorial rectangle satisfying $\mu(-1|R) \le \frac{1}{2}-\frac{1}{6d}$. 
By definition of the discrepancy,  
there is a combinatorial rectangle $R$, such that
\begin{equation}
\label{2}
|\sum_{(i,j)\in R} \mu(i,j)A_{i,j} | \ge \frac{1}{d}.
\end{equation}
Observe that we can assume that
$$
\sum_{(i,j)\in R} \mu(i,j)A_{i,j}  \ge \frac{1}{3d}.
$$
Otherwise, the sum in Equation~(\ref{2}) is negative, and since $\mu$ is uniformly-balanced
$$
\sum_{(i,j)\in \bar{R}} \mu(i,j)A_{i,j}  \ge \frac{1}{d},
$$
where $\bar{R}$ is the complement of $R$. But $\bar{R}$ can be partitioned into
three combinatorial rectangles, and thus there is a rectangle $R'$ such that
$$
\sum_{(i,j)\in R'} \mu(i,j)A_{i,j}  \ge \frac{1}{3d}.
$$

Now, $\sum_{(i,j)\in R} \mu(i,j)A_{i,j} = \mu(1,R)-\mu(-1,R)$, and $\mu(R) = \mu(1,R)+\mu(-1,R)$. 
Therefore,
\begin{eqnarray*}
\mu(-1,R) &=& \frac{1}{2}\mu(R)-\frac{1}{2}\sum_{(i,j)\in R} \mu(i,j)A_{i,j} \\
&\le&  \frac{1}{2}\mu(R)-\frac{1}{6d}\\
&=& (\frac{1}{2}-\frac{1}{6d\mu(R)})\mu(R)\\
&\le& (\frac{1}{2}-\frac{1}{6d})\mu(R).
\end{eqnarray*}
This concludes the proof, as obviously $\mu(R) \ge \frac{1}{d}$.
\end{proof}

\section*{Acknowledgements}

I thank Troy Lee and Michal Parnas for their help in writing this manuscript.


\bibliographystyle{plain}
\bibliography{../complexity}


\end{document}